\documentclass[pre,preprint]{revtex4}
\usepackage{color}

\newcommand{\beq}{\begin{equation}}
\newcommand{\eeq}{\end{equation}}
\newcommand{\eel}[1]{\label{#1}\end{equation}}
\newcommand{\bea}{\begin{eqnarray}}
\newcommand{\eea}{\end{eqnarray}}
\newcommand{\eeal}[1]{\label{#1}\end{eqnarray}}
\newcommand{\beac}{\begin{equation}\begin{array}{rcl}}
\newcommand{\eeacn}[1]{\end{array}\label{#1}\end{equation}}
\newcommand{\beqs}[1]{\begin{equation}\label{#1}\begin{split}}
\newcommand{\eeqs}{\end{split}\begin{equation}}
\newcommand{\non}{\nonumber}

\newcommand\opc[1]{{\cal #1}}

\def\ee{\end{equation}}
\def\be{\begin{equation}}
\def\H{{\cal{H}}}

\newcommand\St{$\cal{S}$}
\newcommand\We{$\cal{W}$}

\begin{document}
\preprint{JCP A7.11.239}
\title{Quasi-equilibrium states in thermotropic liquid crystals studied by multiple quantum NMR} 
\author{L. Buljubasich, G. A. Monti and R.H. Acosta} 
\affiliation{LANAIS - CONICET - Facultad de Matem\'atica, Astronom\'{\i}a y F\'{\i}sica, Universidad Nacional de C\'ordoba, Ciudad Universitaria, X5016LAE - C\'ordoba, Argentina}
\author{C.J. Bonin, C.E. Gonz\' alez and  R.C. Zamar}  \email{zamar@famaf.unc.edu.ar} 
\affiliation{ Facultad de Matem\'atica, Astronom\'{\i}a y F\'{\i}sica,
Universidad Nacional de C\'ordoba - Ciudad Universitaria, X5016LAE - C\'ordoba, Argentina}


\begin{abstract}
Previous work showed that by means of the Jeener-Broekaert (JB) experiment, two quasi-equilibrium states can be selectively prepared in the proton spin system of thermotropic nematic liquid crystals (LC)  in a strong magnetic field. 
The similarity of the experimental results obtained in a variety of LC in a broad Larmor frequency range, with crystal hydrates, supports the assumption that also in LC the two spin reservoirs into which the Zeeman order is transferred, originate in the dipolar energy and that they are associated with a separation in energy scales: a constant of motion related to the stronger dipolar interactions ($\cal{S}$), and a second one ($\cal{W}$) corresponding to the secular part of the weaker dipolar interactions  with regard to the Zeeman and the strong dipolar part. 
We study the nature of these quasiinvariants in nematic 5CB (4'-pentyl-4-biphenyl-carbonitrile) and measure their relaxation times by encoding the multiple quantum coherences of the states following the JB pulse pair on two orthogonal bases, Z and X. The experiments were also performed in powder adamantane at 301 K which is used as a reference compound having only one dipolar quasiinvariant. 
We show that the evolution of the quantum states during the build up of the quasi-equilibrium state in 5CB prepared under the $\cal{S}$ condition is similar to the case of powder adamantane and that their quasi-equilibrium density operators have the same tensor structure. 
In contrast, the second constant of motion, whose explicit operator form is not known, involves a richer composition of multiple quantum coherences on the X basis of even order, in consistency with the truncation inherent in its definition. We exploited the exclusive presence coherences $\pm$4, $\pm$6, $\pm$8, besides 0 and $\pm$2 under the $\cal{W}$ condition to measure the spin-lattice relaxation time $T_{\cal{W}}$ accurately, so avoiding experimental difficulties that usually impair dipolar order relaxation measurement such as Zeeman contamination at high fields, and also superposition of the different quasiinvariants. 
This  procedure opens the possibility of measuring the spin-lattice relaxation of a quasiinvariant independent of the Zeeman and $\cal S$ reservoirs, so incorporating a new  relaxation parameter useful for studying the complex molecular dynamics in mesophases.
 In fact, we report the first measurement of $T_{\cal{W}}$ in a liquid crystal at high magnetic fields. The comparison of the obtained value with the one corresponding to a lower field (16 MHz) points out that the relaxation of the  \We-order strongly depends on the intensity of the external magnetic field, similarly to the case of the \St $\;$ reservoir, indicating that the relaxation of the $\cal{W}$-quasiinvariant is also governed by the cooperative molecular motions. 
\end{abstract}
\maketitle

\section{Introduction}\label{introduccion}

Due to the high degree of orientational order, thermotropic liquid crystals (LC) present strong residual  dipole-dipole couplings between proton spins belonging to the same molecule, while the rapid translational diffusion
 averages the intermolecular dipole coupling to zero \cite{pines88,baum86}.  LC molecules can thus be regarded as separate clusters of dipole coupled spins, where it is possible to prepare dipolar ordered states, besides the usual Zeeman order.
 
By means of the Jeener-Broekaert (JB) radiofrquency pulse sequence \cite{jeene67} ($90^o_x-t_{12}-45^o_y-\tau-45^o_y-t$), it is possible to transfer the initial equilibrium Zeeman order into the dipolar reservoir and to observe the created dipolar ordered state. During the preparation time $t_{12}$, multi-spin single-quantum coherences develop due to the evolution under the dipolar interaction and the second $45^o_y$ pulse transforms part of the coherences just created into multi-spin order  \cite{baum85,walls,cho05}. The quantum state which the spin system is brougth into by the first two pulses evolves during $\tau$, becoming a quasi-equilibrium state over  $\tau$ greater than  $T_2^*$ (the characteristic decay time of the NMR signal, typically, few hundreds of microseconds). The third pulse converts back the dipolar order into observable single quantum coherence, sometimes called `dipolar echo'. 

Two different  quasi-equilibrium states can be selectively prepared in LC in the nematic phase by suitably selecting the preparation time $t_{12}$ of the JB sequence \cite{schmie82,interpar}. It was found that the attributes of the  dipolar echoes  in nematic 5CB (4'-pentyl-4-biphenyl-carbonitrile), in the alkyl deuterated 5BC$_{d11}$ and in PAA$_{d6}$  at 16 and 27 MHz, as well as 300 MHz \cite{interpar,nosphysB,tesismen}, are similar to those of the `intrapair' and `interpair' quasiinvariants measured in crystal hydrates \cite{dumon94,kelle88,eisen78}. That is, both the dipolar echo amplitudes and shapes depend on $t_{12}$ \cite{interpar,tesismen}.  At short preparation times the dipolar echo is proportional to the time derivative of the Zeeman signal (FID), indicating that this state can be represented by an operator similar to the secular high-field dipolar Hamiltonian. The signal shape changes drastically with increasing $t_{12}$ revealing the occurrence of a second quasiinvariant. When the system is prepared in each of these states, the relaxation to the equilibrium with the lattice is characterized by a single exponential. The signals are symmetric with regard to  $t_{12}$ and the observation time $t$, which is consistent with the quasi-equilibrium form of the density operator as a combination of  two commuting, orthogonal operators  \cite{interpar}.

Nematic LC and crystal hydrates also have in common the fact that each spin interacts more strongly with one neighbour than with any other spin, which is reflected in the characteristic doublet shape of their spectra. Physically, as was discussed for solids \cite{jensen95,lefmann94}, this feature indicates that the interaction of a spin with its adjacent surroundings is strong enough to establish a coherent response in spite of the broadening effects due to more distant spins. However, a model of dilute pairs, valid for crystal hydrates, cannot be generally used to represent the spin dynamics in LC in the whole timescale of the experiment because of the proton distribution in LC molecules. Similarities between the experimental results in the cyanobiphenyls, PAA$_{d6}$, and the crystal hydrates, supports the assumption that a general classification of the dipolar interactions into stronger and weaker  can be introduced to describe the quasiinvariants in 5CB and other similar LC.  Accordingly we describe the dipolar energy as a sum of a strong  contribution (${\cal{S}}$) and a weaker one (${\cal{W}}$), the latter truncated to retain its secular part with regard to the Zeeman and ${\cal{S}}$ contributions \cite{interpar}. The partition of the spin-spin energy into two separate contributions like this, generates two constants of motion (quasiinvariants) of very different character, associated with two different timescales of the spin dynamics.
Due to truncation, an operator form of the  \We $\;$ quasiinvariant, for a general ensemble of dipole coupled spins is not known. Therefore, NMR methods that allow unfolding the multiple quantum (MQ) content of the state after the JB preparation pulse  can be used to get a deeper insight on the nature of the quasiinvariants.

 MQ NMR techniques in solids and LC were generally used to study the size of localized spin clusters, to probe the dynamics of many body spin systems, and more recently for the study of decoherence processes \cite{baum86,cho05,cho06,warren80,kroj07,vBeek05}. In all these experiments the starting condition generally was the thermodynamic equilibrium in a strong external magnetic field (Zeeman order).  The dipolar ordered state has  also been considered as the initial state for numerical calculations of MQ experiments \cite{furman05,doron06,doron07}. 
A method for encoding the coherence numbers of the dipolar-ordered state in a basis (X) orthogonal to the Zeeman (Z) basis (in the rotating frame) in solids was presented by H. Cho {\em et al.} \cite{cho03}. In that experiment, the dipolar ordered state created by a JB pulse pair is encoded into zero and double quantum coherence, while the Zeeman order is encoded into single quantum coherence on the X basis. By using this technique, the evolution of the multiple quantum coherences in the initial regime on the two bases was measured in a single crystal of calcium fluoride oriented along the [110] and [100] directions with respect to the external magnetic field.
The neat separation of the coherence terms contained in the dipolar ordered state was useful to characterize the state  and to measure  the dipolar order relaxation time accurately. 

The spin-lattice relaxation of the dipolar energy is very sensitive to the slow collective molecular motion typical of the mesophases,  namely the order director fluctuations (ODF). Contrarily to Zeeman order where this mechanism dominates the relaxation in the range of low frequencies, typically from few KHz to hundreds of  KHz, the ODF have a high relative weight in the dipolar order relaxation even within the MHz Larmor frequency range \cite{jcp98}. Furthermore, the spin-lattice relaxation times  of the two quasiinvariants prepared with the JB pulse pair are noticeably different and in some compounds, they have also different temperature behavior. The relaxation of the \We-order is much more efficient than the ${\cal{S}}$ one through the whole nematic temperature range \cite{interpar,nosphysB,jcp05}. Measurement of additional relaxation parameters other than the usual Zeeman relaxation time  can help to disentangle the relevant  spectral densities of the  complex  molecular motions in LCs \cite{dongbook}.
However, dipolar relaxation experiments generally present certain degree of difficulty due to possible contamination with Zeeman magnetization of the signal observed in the `dipolar channel' \cite{yousef}. 
A more subtle difficulty is related to selectively transferring the Zeeman order to the \We $\;$ reservoir, because the range of $t_{12}$ where this condition is met is very narrow \cite{interpar}. A superposition of both kinds of quasiinvariants due to inaccurate setting of this  condition may spoil the  relaxation time measurement. In fact, to our knowledge, no measurement of this parameter in LCs at high magnetic fields has been reported. 

 In this work we follow the ideas presented in reference \cite{cho03} to monitor the proton spin dynamics in the quantum states obtained after the JB pulse pair in nematic 5CB at 300 MHz. Preparation times corresponding to the two quasiinvariants are used and the spin-lattice relaxation for each one is measured. 
Rotating the state around an axis orthogonal to Z allows encoding MQ coherences which reflect 
the number of multiply connected spins in the quasiinvariant states.
Furthermore we explore the nature of the states prepared with the JB pulse pair and study the creation of the two quasiinvariants. The experiments were also performed in powder adamantane which is used as a reference compound having only one dipolar quasiinvariant.  We show that the order associated with the strong dipolar energy resembles the behavior observed in ordinary solids, as expected. In the course of the experiments the presence of only even coherence orders greater than two on the X basis were found for the ${\cal{W}}$ reservoir.  We interpret this fact as a confirmation that both quasi-invariants originate in the dipolar spin-spin coupling and as an evidence of the multiply connected nature of the \We $\;$ quasiinvariant. The occurrence of different coherence numbers of the ordered state associated with the weaker couplings are used to avoid the different sources of contamination mentioned above; by using this method we could monitor unambiguously the decay of the ${\cal{W}}$ quasiinvariant towards the equilibrium. In fact, we report the first measurement of the relaxation time $T_{\cal{W}}$  in a liquid crystal at high magnetic fields. 

\section{Constants of the motion in liquid crystals}
 
In a strong magnetic field {\bf B$_o$},  the interaction energy of a system of nuclear dipole coupled spins  can be represented by the secular part of the dipolar Hamiltonian, which, in units of $\hbar$ is
\be\label{eq:rho_2_0}
\opc{H}_D^o = \sqrt{6}\sum_{i<j}D_{ij}T_{20}^{ij},
\label{hamdip}  
\ee
where the sum runs over all the interacting protons. 
In LC, the  dipolar couplings $D_{ij}$  are  averaged  over the fast molecular  motions \cite{segno06,dongbook}. 
\begin{equation}
D_{ij}\equiv \left< \frac{\mu_{o}\gamma^{2}\hbar}{4\pi}\left( \frac{1-3cos^{2}\theta_{ij}}{2r_{ij}^{3}}\right)\right> , 
\end{equation}
where $r_{ij}$ is the internuclear distance between  nuclei $i$ and $j$ and $\theta_{ij}$ is the angle between the internuclear vector and the magnetic field \textbf{B$_o$}. Since the intermolecular dipolar interactions average to zero \cite{pines88}, indices $i,j$ run only within each molecule. 
$T_{20}^{ij}$ is the zero component of a normalized irreducible spherical tensor of rank two, which in terms of the individual spin angular momentum operators is \cite{abragam, abragol}
\begin{equation}\label{t20}
T_{20}^{ij}=\frac{1}{\sqrt{6}}\left[ 2I_{z}^{i}I_{z}^{j}-\frac{1}{2}\left(I_{+}^{i}I_{-}^{j}+I_{-}^{i}I_{+}^{j}\right)\right].
\end{equation} 

The basic characteristic which 5CB shares with other similar LCs and crystal hydrates is the occurrence of a separation of energy scales within the spin interactions \cite{acoplesdipolares}, which also implies the occurrence of two timescales in the spin dynamics. Then, at high magnetic field it is justified to define
\be
||\H_Z|| \gg||\H_{\cal{S}}|| \gg ||\H_{\cal{W}}||, \label{jerarquia}
\ee
where the subscript $\cal{S}$ stands for the subset of stronger interactions,  $\cal{W}$ for the weaker ones and
$\mathcal{H}_Z=\omega_0I_Z$, is the usual Zeeman energy where $\omega_o$  the Larmor frequency. The secular dipolar Hamiltonian then is 
$
\opc{H}_D^o = \opc{H}_{\cal{S}} + \opc{H}_{\cal{W}} \,, 
$
with
\be
\H_{\cal{S}}= \frac{\sqrt{6}}{2}\sum_{(i,j) \; \in \cal{S}} D_{ij} {\bf T}_{20}^{i j}\quad  {\rm and} 
\quad \opc{H}_{\cal{W}}= \frac{\sqrt{6}}{2}\sum_{(i,j) \; \in \cal{W}} D_{ij} {\bf T}_{20}^{i j},\label{HSHW}
\ee
 where the subscript $(i,j)$ was used to emphasize that pairs rather than individual
spins belong to \St $\;$ and \We.

In the common basis of $\H_{\cal{S}}$ and $\H_Z$, the weaker Hamiltonian can be written as the sum of a diagonal (in blocks) and a nondiagonal term, 
\be
\H_{\cal W}=\H^{\rm d}_{\cal W}+\H^{\rm nd}_{\cal W},
\label{Hw}
\ee
and the hierarchy of Eq.(\ref{jerarquia}) allows truncating this Hamiltonian by eliminating the nondiagonal term, so keeping its secular part with regard to $\H_{\cal{S}}$ and $\H_Z$. 
It is worth to notice that, because of the truncation inherent in its definition, $\H^{\rm d}_{\cal W}$ does not preserve the bilinear form of  Eq. (\ref{hamdip}), that is, it may have a more complex structure of multiple-spin nature, which we probe in this work. According to these definitions, the total energy of the spin system can be considered  a sum of three orthogonal constants of motion  
\be
\opc{H} = \opc{H}_{Z} + \H_{\cal{S}} + \H^{\rm d}_{\cal{W}} \:,
\label{energy}
\ee
which satisfy 
\begin{eqnarray}
\left[\opc{H}_Z,\opc{H}^{\rm d}_{\cal{W}}\right] = 0, \quad 
&\left[\opc{H}_{\cal{S}},\opc{H}^{\rm d}_{\cal{W}}\right] = 0, \\
\left[\opc{H}_Z,\opc{H}_{\cal{S}}\right]=0. \non 
\end{eqnarray}

  A closed expression for $\H^{\rm d}_{\cal{W}}$ was given only for the case of weakly interacting spin pairs treated as a spin-1 system \cite{kelle88}, which we reproduce in Appendix A for illustration. There, the operator of Eq.(\ref{HinterX}) includes products up to four components of the spin angular moment. 

In summary, we assume that the dipolar Hamiltonian admits a partition into two categories of dipolar couplings, such that the non-secular part of the weak dipolar terms (with respect to the Zeeman and strong dipolar terms) does not influence the evolution of the spin coherence during the relevant experimental timescale. This is analog to the perturbative approach  used in truncating the complete dipolar Hamiltonian with regard to the Zeeman energy at high magnetic fields \cite{abragam}. 

By applying the preparation pulses of the JB sequence to a state of Zeeman order $\rho_0 \propto \mathcal{I}-\beta_0\opc{H}_Z$   ($\beta_0=1/kT$ the inverse equilibrium temperature of the lattice and assuming high temperature) the proton spin system can be brought into a state of dipolar order over a time $\tau$ long enough to allow the off-diagonal matrix elements to decay to zero. That is, the system reaches a state of internal equilibrium which is completely specified by the constants of motion \cite{jeene68}. In a spin system like 5CB, where the spin Hamiltonian can be written in terms of three constants of motion as in Eq.(\ref{energy}), the quasi-equilibrium density operator in the rotating frame has the form \cite{interpar} (we dropped the normalization constant for brevity) 
\begin{equation}\label{dipolar_5cb}
\rho(\tau,t_{12})\approx \mathcal{I}- \beta_{\cal{S}}(t_{12})\opc{H}_{\cal{S}}-\beta_{\cal{W}}(t_{12})\H^{\rm d}_{\cal{W}} \;,
\end{equation}
where $\beta_{\cal{S}}$ and $\beta_{\cal{W}}$ are the inverse $\cal{S}$- and $\cal{W}$-order temperature, respectively. We assumed that $t_{12} \ll T_{1Z}$, and then the Zeeman inverse temperature $\beta_{Z}(t_{12})\simeq 0$.
In an ordinary solid with two constants of motion, like adamantane,
\begin{equation}\label{dipolar_adamantane}
\rho(\tau,t_{12})\approx \mathcal{I}-\beta_{D}(t_{12})\opc{H}_{D}^{o}\;,
\end{equation}
where $\beta_{D}$ is the dipolar inverse spin temperature. 

By adequately setting the preparation time of the JB sequence, $t_{12}$, it is possible to regulate the values of the inverse spin temperatures and to prepare a state of ``\St-order'' with  $\beta_{\cal{W}}(t_{12})=0$  or a state of pure ``\We-order''   with $\beta_{\cal{S}}(t_{12})=0$.

\section{Preparation of ordered states} 

In this section we show the experimental signals from 5CB and powder adamantane after the application of the JB pulse sequence  in order to clarify the very different nature of their spin systems. All the experiments were performed on-resonance at 7 T using a Bruker Avance II spectrometer. A DOTY DSI-703 proton dedicated probe with proton background signal reduction was used. Samples were packed in 4 mm outer diameter ZrO sample holders fitted with Kel-F end caps. The length of the $\pi$/2 pulses was of 2 $\mu$s. Measurements were carried out at a temperature of 301 K for both samples. The on-resonance condition for 5CB corresponds to the on-resonance condition for the isotropic phase. The Zeeman spin-lattice relaxation time is $T_1=630$ ms and $T_1=1.2$ s for 5CB and powder adamantane, respectively.   

 An indication of the occurrence of two  quasiinvariants, besides the Zeeman energy in LC in the nematic phase, is   the variation of the dipolar signal shape with the preparation time $t_{12}$ in the JB pulse sequence \cite{interpar,weitekamp}. Fig.  1(a) shows this behaviour in 5CB. It was shown that these signals are symmetric with respect to the preparation and observation times $t_{12}$ and $t$ \cite{interpar}.
  The $\cal{S}$ condition, which is also the dipolar signal with maximum amplitude, is obtained by setting $t_{12}$ equal to the time corresponding to the maximum derivative of the Zeeman signal (dotted curve in Fig. 1(a)). 
As a consequence of the symmetry in $t_{12}$ and $t$ the dipolar signals at a time $t$ can also be interpreted as a measure of the dipolar order at the corresponding preparation time $t_{12}=t$ \cite{interpar}. Therefore, pure $\cal{W}$-order can be selected by choosing $t_{12}$ so that the $\cal{S}$-signal crosses through zero. 
The solid and dashed curves of Fig. 1(a) are the NMR signals of the states of $\cal{S}$- and $\cal{W}$-order in 5CB. At the temperature of this experiment they correspond to $t_{12}=28$ $\mu$s and $t_{12}=70$ $\mu$s respectively, and $\tau$ was set to 2 ms.  
It is worth to notice that a small shift from such condition produces a rapid increase of the $\cal{S}$ component; in practice, an error of a few microseconds in setting $t_{12}$ may be a source of ``contamination'' of the $\cal{W}$-signal with an $\cal{S}$ component. On the other hand, the situation of maximum $\cal{S}$-order coincides with a small amount of $\cal{W}$ order (about 4\%). In section \ref{T1D} we use the relaxation times  to show that the $\cal{W}$ contamination on the $\cal{S}$-order is negligible and that a careful setting of the $\cal{W}$ condition leads to a non-contaminated signal.  

The Zeeman and dipolar signals of powder adamantane, for different values of $t_{12}$ are illustrated in Figure 1(b), where $\tau$ was also set to 2 ms. The dipolar signal of maximum amplitude corresponds to $t_{12}$= 40 $\mu$s that is also the time at which the time derivative of the FID takes its maximum value. It can be observed that all the dipolar signals have the same shape. All the maxima  occur at the same time and all of them cross through zero at the time corresponding to the minimum of the FID at $t=107$ $\mu$s.
These features are characteristic of the presence of a single dipolar quasiinvariant resulting in a proportionality between all the dipolar signals and the time derivative of the FID \cite{jeene67,abragol}.

\section{Multiple quantum coherence in X and Z bases}

In order to estimate the state of the $^1$H spin system in LC when prepared in 
quasi-equilibrium states,  as well as to follow the coherent spin dynamics in the transient prior to the establishment of the quasi-equilibrium states,  we used a version of the  pulse experiment proposed in reference \cite{cho03}. 
 A schematic diagram of the pulse sequence is shown in Fig.  2. The first two pulses are the JB preparation pulses \cite{jeene67}. Varying $\eta$ systematically in succesive experiments allows encoding in the Z basis. The following two pulses $90^o_{(\phi + \pi/2)}-\epsilon-90^o_{(-y)}$ encode the coherence numbers of the quantum state at time $\tau$ in the X basis when varying $\phi$ systematically in succesive experiments. The time $\delta$ before the $45^o_{y}$ read pulse was set long enough to allow undesired signals to decay (see section \ref{senhal}): $\delta=1$ms for 5CB and $\delta=500$ $\mu$s for adamantane. In this work we used a time interval $\epsilon = 1$ $\mu$s, much smaller than the dipolar periods instead of the 48-pulse sequence \cite{cory90} used in reference \cite{cho03}. By using a small $\epsilon$ we assume that no significant evolution of the state of the spin system under the dipolar Hamiltonian occurs during this period, so that the effect of the third and fourth pulses is essentially a rotation around the $x$-axis by an angle $\phi$. By  changing $\phi$ and $\eta$ independently it is possible to  encode and observe simultaneously correlations between the coherences in the X and Z bases respectively. 

The state of the spin system can then be traced  for short times  $\tau$ during the coherent evolution towards the quasi-equilibrium states. By studying the relative amplitudes of the different coherences in the asymptotic regime, in both bases, it is possible to probe into the nature of the  quasiinvariant states that can be prepared by suitably adjusting the preparation time $t_{12}$.
Finally, their spin-lattice relaxation times can be measured by recording the time evolution of the coherence amplitudes when $\tau$ is varied along a larger time scale. 

 For coherence encoding up to order $m$ on both bases, the phases $\phi$ and $\eta$ were incremented from 0 to 2$\pi$ with $\Delta \phi=\Delta \eta = \pi/m$. The coherence encoded signals for each value of $\phi$ and $\eta$ were integrated around the maximum intensity. 
Data were Fourier transformed with respect to both phases to yield the coherence numbers.

\subsection{Description of the encoded dipolar signal}\label{senhal}

If we assume that the state prepared at time $\tau$ corresponds either to Eq. (\ref{dipolar_adamantane}) for powder adamantane or to \St-order in 5CB (Eq. (\ref{dipolar_5cb})), the observed signal at time $t$ after the readout pulse (5$^{th}$ pulse in Fig. 2), neglecting spin-lattice relaxation effects, can be written as 

\begin{eqnarray}  \label{signal}
&&S(\phi,\eta,\delta,t)\equiv \left<I_y \right> \\
&&\propto \beta_D e^{-i0\eta}\left[\left(\frac{3}{2}\cos^2{\phi}-\frac{1}{2}\right)G'(t) -  \sin^2{\phi} {\tt Tr}\{I_y U H_{2}U^{\dagger}\} \right],\non
\end{eqnarray}
where $\beta_D$ represents the inverse dipolar temperature of the prepared order, 
$$G'(t)\equiv{\tt Tr}\{I_y e^{-i \opc{H}_D^0 t}[I_y,i \H'] e^{i \opc{H}_D^0 t} \}\;,$$
with $\H'$ equal to $\H_D^0$ for powder adamantane or to $\H_{\cal S}$ for 5CB. Therefore, $G'(t)$ practically coincides with the time derivative of the FID signal after a 90$^o$ pulse. In the second term $U\equiv e^{-i\H_D^0 t}e^{i\frac{\pi}{4}I_y}e^{-i\H_D^0 \delta}$, and 
$$H_2\equiv\sqrt{\frac{3}{8}}\sum_{(i,j)}D_{ij}\left(T_{22}^{ij}+T_{2-2}^{ij} \right),$$  where $i,j \in $ \St $\;$ for 5CB. We have also assumed that $\epsilon=0$ and that the pulses were applied on resonance. The factor $e^{-i0\eta}$ points out that only zero quantum coherence is encoded on the Z basis once the quasi-equilibrium is established.

The first term of Eq. (\ref{signal}) is proportional to the dipolar signal \cite{jeene67,interpar} modulated by a $\phi$ dependent factor. The second term involves  an additional dependence on  $\delta$ through the time evolution operator $U$. This  contribution attenuates with $\delta$, for fixed $t$, with the characteristic decay time of the double  quantum coherence. In consequence, for sufficiently long $\delta$ we can expect the observed signal to be represented by the first term only.
A representative signal for \St-order in 5CB, with $\tau=300\;\mu$s, $\epsilon=1 \;\mu$s and $\delta=1$ ms is shown after the read pulse in Fig.  2. 
Under these conditions, the Fourier transform of  the signal $S(\phi,\eta,\delta,t)$ (Eq. (\ref{signal}))  respect to  $\eta$ and $\phi$ will present only zero quantum coherence  in the Z basis, and 0 and $\pm$2 in the X basis.  Accordingly, we can expect a ratio $R_{20}\simeq $ 1.5 between double and zero quantum coherence amplitudes in the X basis \cite{cho03} in our experiment.


\subsection{Initial spin dynamics after JB pulses} 

In this section we study the coherent  dynamics of the spin state following the JB sequence, prior attaining the quasi-equilibrium states. We present the first results on the evolution of MQ coherences during the creation of two different ordered states in LCs. Encoding was performed up to coherence order m=8. Additionally, in the case of the \We $\;$ reservoir, encoding up to coherence order m=16 was carried out. Coherence orders higher than order 8 were seen to fall well below the noise level on all experiments.

Figure 3 shows the 2D encoded measurement of the coherences on the Z and X bases in powder adamantane and 5CB.
On a very short time scale of the evolution $\tau$, coherence numbers  $x$=0 and $x=\pm$2 are observed while on the Z basis, coherence orders up to  $\pm3$ (odd and even) can be detected both in adamantane (Fig. 3(a)) and 5CB prepared in the \St $\;$condition (Fig. 3(b)).
 The largest intensity pixels correspond to coherence numbers $z = 0$ and $x = \pm$2. Over a time $\tau\approx 300$ $\mu$s, coherences on the Z basis have decayed and only terms with $z = 0$ and $x = 0, \pm$2 remain, similarly to the case of calcium fluoride \cite{cho03}.
Figure 3(c) shows the results obtained for \We-order. Initially, a broader distribution of coherence numbers emerges in the X basis. In this case, also coherences of even order, higher than $\pm$2 remain on the X basis and $z = 0$ for larger $\tau$. This feature is an evidence of the more complex structure of the \We-order. 

The time behavior of the different coherences is more clearly appreciated by taking  a projection over each axis, as shown in Fig. 4. Evolution of the coherence terms for adamantane and \St-order in 5CB appear very similar, as seen in Figs. 4(a) and 4(b). Zero quantum terms remain nearly constant on the Z basis on this time window, while single quantum terms attenuate within 200 $\mu$s and show a peak at 65 $\mu$s in adamantane and 30 $\mu$s in 5CB.  Double quantum coherence (DQC) in the Z basis starts from a maximum and decay to zero in both cases showing similar profiles to those reported in ref. \cite{emid}. 
On the X basis only zero and DQC have a significant amplitude; after an initial oscillation a plateau consistent with the establishment of a quasi-equilibrium state is reached at times around 150-200 $\mu$s.
Also a component of coherence number $x=\pm 4$ with an amplitude about two orders of magnitude smaller than the other signals was detected in 5CB in the \St $\;$condition (this component is not shown in Fig. 4b). The origin of this signal is analyzed in the next section. 

Evolution of the states prepared in the \We $\;$condition present several differences with respect to the \St$\;$ case. Zero quantum coherence encoded on the Z basis shows a higher slope than the one observed for the other two cases and the single quantum coherence shows a marked oscillation prior to its decay as shown in Fig. 4(c). 
Also noticeable differences are seen in the  early behaviour of coherences encoded on the X basis. Not only are oscillations more evident for DQC but a crossover on the intensities with zero quantum coherence occurs for $\tau \approx$ 180 $\mu$s. Higher order terms of even order (only x = $\pm$4 is plotted  in Fig. 4(c)) seem to reach a plateau over 500 $\mu$s.  However, it is interesting to study the coherence evolution on a longer timescale.  
We therefore scanned their behaviour with the evolution time $\tau$ in small steps up to 5 ms and plotted them in Figure 4d. Coherences 6 and 8 keep growing during the first millisecond, developing a parallel time behaviour which is opposite to the trend of the other terms (0, 2 and 4). In fact, a quasi-equilibrium state, where all the peaks evolve with the same rate, is only attained over 2ms. That is, the evolution of the coherent state towards the \We $\;$ quasi-equilibrium is significantly slower than the \St $\;$ one.


Figures 5(a) and 5(b) show the normalized coherence spectrum of adamantane and 5CB prepared in the dipolar \St-order, for an evolution time of 300 $\mu$s. For this timing  the coherences  have reached a plateau according to the effective formation of a dipolar ordered state. As seen in the figures, these spectra are similar.

By measuring a ratio $R_{20}\simeq 1.5$ between double quantum and zero quantum coherence amplitudes (on the X basis) after preparing the dipolar state both by adiabatic demagnetization in the rotating frame (ADRF) and by the JB sequence, and using adiabatic remagnetization (ARRF) to convert back into observable magnetization, H. Cho {\it et al.} \cite{cho03} confirmed that in calcium fluoride the density operator corresponding to the quasi-equilibrium state reached after preparation is that of Eq.(\ref{dipolar_adamantane}). In this work we measure the ratio $R_{20}$ yielded by the experiment of Fig. 2 for $\tau$=300$\mu$s. We obtained $R_{20}\approx1.6$ for powder adamantane and $R_{20}\approx 1.7$ for the \St $\;$ dipolar order in 5CB, as can be noticed from Figs. 5(a) and 5(b). These values are near the quotient 1.5 expected from Eq.(\ref{signal}). 

The coherence content is very different for \We-order, as shown in Fig. 5(c) (for $\tau$=2 ms). Not just the zero quantum coherence presents a higher intensity than the DQC, but  also a significant $\pm 4$ projection is clearly  appreciated, and smaller amplitude contributions of $\pm6$ and  $\pm8$ order can be detected. The fact that the projection on the X basis of the \We-state has coherences of even order higher than 2 indicates that the tensor structure of $\H^{\rm d}_{\cal W}$ differs from a bilinear form. 

\section{Relaxation measurements on the X basis}\label{T1D}

\subsection{Zeeman contamination}

Separation of different coherence numbers in the Z and X basis allows measuring dipolar relaxation times avoiding possible contamination of data with Zeeman magnetization \cite{yousef}, as proposed by H. Cho {\it et al.} \cite{cho03}. 
Since the Zeeman term is encoded into single quantum coherence on the X basis, while the dipolar contribution is encoded into zero and even order coherences, a clean measurement of the dipolar relaxation may be accomplished by recording the evolution with long $\tau$ of the different coherence numbers on the X basis. 
Figures 6(a) and 6(b) show that the amplitude of both coherence $x = 0$ and $x=\pm2$ attenuate as single exponentials with a characteristic time $ T_{1D}$ = (665 $ \pm$ 30)ms in powder adamantane and $T_{\cal{S}}$=(310 $\pm$ 13)ms for the $\cal{S}$-order in 5CB.

The method is still more useful to overcome important difficulties that arise in the $T_{\cal{W}}$ experiment. Since the signal amplitudes are smaller than the $\cal{S}$ case, Zeeman contamination usually impairs $T_{\cal{W}}$ measurement, specially at high fields. Figure 6(c) shows the evolution with $\tau$ of all the coherences observed for the $\cal{W}$ condition. 
The straight lines shown in Figs. 6 (a),(b) and (c) are the fittings corresponding to the evolution of the coherence $x = 0$. It is worth to notice that all data sets present single-exponential behavior. Their characteristic decay times are shown in Fig. 7. For adamantane and 5CB in the 
$\cal{S}$-order, the values obtained for $x = 0$ and $x=\pm2$ agree within 5\% and for 5CB with $\cal {W}$-order the results obtained for $x = 0, \pm2, \pm4, \pm6$ and $\pm8$ agree within 10\%, yielding a mean value $T_{\cal{W}}$ = (126 $\pm$ 15)ms. We can then assert that the relaxation time is a signature of the different ordered states evolving under each coherence number, and can therefore be used to determine the nature of each coherence number present in the different quasiinvariants.

\subsection{$\cal {W}$ contamination on the determination of $T_{\cal {S}}$}

As mentioned before, a small component of $x = \pm 4$ could be observed on the $\cal{S}$ spectrum, which in principle does not agree with a state represented by Eq. (\ref{dipolar_5cb}) with  $\beta_{\cal {W}}$=0 (pure $\cal{S}$ state). Nevertheless, a careful inspection of the evolution of this coherence (see inset in Fig. 6(b)) shows that in fact the associated relaxation time (140 $\pm$ 37)ms clearly coincides with $ T_{\cal {W}}$ instead of $ T_{\cal{S}}$. Thus, this small component can be explained as a contamination from $\cal {W}$-order. Indeed, from Fig. 1(a) we can see that when preparing the $\cal{S}$-order with $t _{12}$ = 28 $\mu$s, a small amount of $\cal {W}$-order has already built up. Although this contamination is also present for $x = 0$ and $x=\pm2$, it is negligibly small as can be inferred from the single-exponential decays.

\subsection{$\cal{S}$ contamination on the determination of $T_{\cal{W}}$}

The very careful selection of the time in which $\cal {W}$-order is created in order to render a pure state is proved in Fig. 7. If this were not the case, contamination from the $\cal{S}$-dipolar order would have drastically influenced the relaxation times for coherence numbers $x = 0$ and $\pm 2$. This was tested by performing an experiment in which $t_{12}$ was intentionally mismatched by a few microseconds. In this case the relaxation times obtained for $x = 0$ and $\pm 2$ differed from those obtained for coherence numbers  $\pm 4$ and greater. Since the latter are exclusively associated with the $\cal {W}$-order, they become a signature of this kind of state. This fact can be exploited to measure $T_{\cal {W}}$ accurately. 

\section{Discussion}

Our experimental results show that the quasi-equilibrium state in powder adamantane is well represented by the density operator of Eq.(\ref{dipolar_adamantane}), as in CaF$_2$. In the case of 5CB, the quasi-equilibrium states are well represented by Eq. (\ref{dipolar_5cb}), where the coefficients $\beta_{\cal S}$ and $\beta_{\cal W}$ characterize the degree of order transferred by the JB sequence from the Zeeman reservoir to the ${\cal S}$- and ${\cal W}$ reservoirs. Under the \St$\;$condition, the evolution of the different coherences and the ratio $R_{20}$ are similar to adamantane, in agreement with $\opc{H}_{\cal{S}}$ having the bilinear form of Eq.(\ref{HSHW}), where the sum runs over a subset of strong  pairwise interactions of the 5CB molecule.

In contrast, the state prepared under the ${\cal W}$ condition encodes into higher coherence numbers on the X basis than the \St $\;$ state. In this experiment we detected even coherences up to order eight. This fact indicates that the ${\cal W}$ state must be represented by a density operator more complex than the bilinear one of the ${\cal S}$ state because the occurrence of coherence orders higher than two require products of more than two operators of spin angular momentum. Also, the fact that only even coherence numbers are observed on the X basis, rules out the possibility of relating the \We-order with other kind of spin interactions like chemical shift effects. These experimental features then suggest that the two quasinvariants derive from the dipolar energy.

Formally, one should expect that the time evolution operator of a system having N degrees of freedom can be written as the sum of at least $2^{\rm N}$  terms, each of which commutes with the total Hamiltonian. However, one is generally able to distinguish only a small number of invariants. 
For example in an ordinary solid like CaF$_2$, the Zeeman and the secular dipolar energy are the only observed quasi-invariants, for all orientations of the external magnetic field \cite{jeene67,cho03}. J.D. Walls and Y. Lin \cite{walls} presented a method for constructing a set of orthogonal constants of motion in terms of products of spin operators, starting from the Zeeman and secular dipolar energy. They found that the projection of the spin state after the JB sequence
 onto operators of multispin character tends to zero when the system (regular linear chains of N spins $\frac{1}{2}$) has enough size: N$> $8 (``thermodynamic limit") \cite{walls,weitekamp}.
 It appears that the evolution of the spin system under the dipolar Hamiltonian during the preparation period cannot efficiently generate multi-spin correlations from Zeeman order.

In other cases, like hydrated salts and LC, the particular symmetry and orientation respect to the external magnetic field allow the couplings with a few neighbours to contribute with a great part of the total dipolar energy. This leads to a splitting of the resonance into two peaks associated with the coherent dynamics of a spin with its neigbours, which is not masked by the incoherent effect  from the multitude of other neighbours (local field), which cause the line width. The separation in energy scales allows the evolution operator to be expressed as a product, and this fact is a condition for the occurrence of the \St $\;$and \We $\;$reservoirs and two different timescales in the spin dynamics. This feature, in the end, is dictated by the topology of the spin distribution in the lattice.

The evolution of the coherent state towards the \We $\;$ quasi-equilibrium is significantly slower than the \St $\;$ one, as can be concluded by comparing Fig. 4b with Fig. 4d, where coherences $x$=0 and $x=\pm$2 on the \St $\;$ condition attain their quasi-equilibrium value within a short period of about 200 $\mu$s. The behavior shown in Fig. 4d is also compatible with the fact that the $z$=0 peak in the \We $\;$ condition, does not reach a plateau within the first 500 $\mu$s  either. The outstanding difference shown by the \We $\;$ state is consistent with the occurrence of multispin correlations involving a great number of spins, which evolve in a longer timescale. The high order peaks on the X basis are a consequence of the multi-spin single-quantum coherences which develop due to the evolution under the dipolar interaction during the preparation period of the JB sequence and that are transformed into multi-spin order by the second JB pulse. In common solids such multi-spin coherences cannot generate during evolution under $\opc{H}_D^o$ probably because of incoherent effects of the multitude of other spins in the lattice (local field).
  
Since the tensor form of $\H^{\rm d}_{\cal W}$ for a spin cluster like 5CB molecule is unknown, it is informative to bring into the discussion the `interpair' Hamiltonian proposed by A. Keller \cite{kelle88} for representing a system of weakly interacting pairs in potassium oxalate monohydrate (POMH), treated as a spin-1 system, and to analyze its projection on the X basis. Though this Hamiltonian corresponds to an ensemble of distant interacting pairs, and hence cannot be used to make any quantitative calculation in 5CB, it serves to illustrate the effects of the rotation of the quantization axis on the coherence content of a truncated dipolar Hamiltonian. 
As shown in Appendix A, such model predicts even order coherences up to $\pm 4$ on the X basis, showing the occurrence of multispin order due to truncation of the dipolar interpair energy. 
The amplitudes of the different coherence order terms on the X basis yielded by  Eq. (\ref{HinterX}) have  a maximum in $x=0$, and decrease for $x=\pm$2 and $x=\pm$4, which is in qualitative agreement with the coherence spectrum of Fig. 5(c). However, the amplitude ratios do not agree with the 5CB experiment, as expected, since the model of weakly-interacting pairs does not reflect the actual complexity of the 5CB dipolar network. 

The occurrence of coherence orders higher than four is also consistent with classifying dipolar couplings into strong and weak. A precise determination of such subsets would demand a whole different strategy. Nevertheless, guided by the description given in \cite{acoplesdipolares} one can conjecture that a reasonable partition which reflects the multiple-spin nature of the \We $\;$ reservoir may assign the dipolar interactions of each spin with its first two neighbours to $\opc{H}_{\cal{S}}$ and the rest to $\opc{H}_{\cal{W}}$. 

Once the multispin nature of the \We $\;$ reservoir has been demonstrated, a next step  towards a formal expression for $\H_{\cal{W}}$  could be a full decomposition of the spin state prepared with the JB pulse pair into spherical tensor components. Such a goal could be achieved through the technique known as ``spherical tensor analysis", recently presented in reference \cite{vBeek05}.

The exponential relaxation observed when the spin system is selectively prepared in each ordered state is consistent with the former view, in which both reservoirs have a many-body character,  with a single `spin temperature' representing the whole molecule in each quasiequilibrium state.
 The fact that all the coherences that characterize the \We $\;$ state relax exponentially with the same decay rate, provides an experimental confirmation that the corresponding $T_{\cal W}$ is the characteristic decay time of an  actual quasiinvariant.  

A result which is very useful for applications is the possibility of measuring  $T_{\cal W}$ in a precise way by following the evolution of the amplitude of the fourth order coherence, which is exclusive of this ordered state.  This  procedure opens the possibility of accurately measuring the spin-lattice relaxation rate of a quasiinvariant independent of the Zeeman- and $\cal S$ reservoirs, so incorporating a new relaxation parameter useful for studying the complex molecular dynamics in mesophases.

It was already shown that neither rotational diffusion nor intermolecular fluctuations mediated by translational self-diffusion play a significant role in the $\cal{S}$ relaxation of PAA$_{d6}$, 5CB and 5CB$_{d11}$ for  $\omega_o$ in the range of decades of MHz. On the contrary, the ODF and the reorientation of the alkyl chain protons (in 5CB) are the mechanisms governing the Larmor frequency dependence of $T_{\cal S}(\omega_o) $ in a similar frequency range \cite{segno06,jcp05}. In all cases, the experimental $T_{\cal S}(\omega_o)$ clearly reflects the well known $\omega_o^{1/2}$ Larmor frequency dependence of the spectral density associated with  the fluctuations of the dipolar energy driven by the ODF. However the high temperature Markovian theory is insufficient to explain the strong influence of the cooperative motions on the $\cal{S}$ relaxation, because of the neglect of correlations between the spin density operator and the lattice density operator in the microscopic timescale, which is an intrinsic hypothesis of the theory. The discrepancy between the experiment and the standard theory is even  more marked for the relaxation of the  \We $\;$ reservoir \cite{segno06}.
In addition, the possible occurrence of ultraslow modes of the ODF with correlation times comparable with the dipolar relaxation times, would demand generalizing the quasi-invariant relaxation theory, using a quantum formalism adequate for the non-Markovian regime \cite{charpentier,jeene90}.

The value  for $T_{\cal W}$ at 300 MHz (126 ms) obtained in this work can be compared with the  previously reported for this parameter  in 5CB at  16 MHz $T_{\cal W}$= 44 ms \cite{nosphysB}.  Our data indicate for the first time that the relaxation of the ${\cal W}$-order in liquid crystals is strongly dependent on the intensity of the external magnetic field just as the ${\cal S}$-order. This suggests that, like in the case of the $\cal{S}$ reservoir \cite{jcp05}, the ODF would be an important source of the strong Larmor frequency dependence of $T_{\cal W}(\omega_o) $ \cite{segno06}. The standard relaxation theory in the high temperature regime predicts $T_{\cal W}\simeq T_{\cal S}$ \cite{segno06} which is a consequence of the semiclassical nature of such  theory, however we observe that $T_{\cal W}\ll T_{\cal S}$ in 5CB, similarly to the case of PAA$d_6$ \cite{interpar}. This fact would then be pointing out the importance of the multiple-spin dynamics in the microscopic timescale. 

Further investigations should be carried out to explore if the intermolecular interactions have any influence on the quasiinvariants created with the JB experiment, especially on the ${\cal W}$-reservoir. The very small dipolar interactions between distant spins in solution and even in gases have been shown to generate macroscopic signals if the spatial symmetry of the sample is broken \cite{richt95,jeene95,chen01,marques05,acosta08,branc08}. These long range couplings could be expected to be more pronounced in a LC due to the restricted molecular diffusion \cite{noack99} and the instrinsic asymmetry on the sample inntroduced by the molecular alignment. 
Also, due to the high long-range molecular correlation of liquid crystals, intermolecular contributions should be expected to contribute to dipolar order relaxation and decoherence.

The state of $\cal{W}$-order can also be used to generalize the studies of spin dynamics in experiments of MQ NMR where the initial state is one of multispin order, instead of the usual one-spin correlated Zeeman order \cite{furman05,doron06,doron07}.

\section{Acknowledgement}
We would like to thank H\'ector H. Segnorile for useful discussions. This work was supported by Secretar\'{\i}a de Ciencia y T\'ecnica from  Universidad Nacional de C\'ordoba, Consejo Nacional de Investigaciones Cient\'{\i}ficas y T\'ecnicas (CONICET), Agencia Nacional de Promoci\'on Cient\'{\i}fica y T\'ecnica (ANPCyT)(Argentina) and the Partner Group with the Max-Planck Institute for Polymer Research, Mainz, Germany. C.J.B.  acknowledges CONICET for financial support.


{\appendix
\section{}
By treating the system of weakly dipolar interacting strong pairs in potassium oxalate monohydrate (POMH) as an ensemble of spin-1 entities,  A. Keller \cite{kelle88} proposed the following expression for the truncated interpair Hamiltonian (in the usual Z basis), which is obtained under the same rule of our $\H^{\rm d}_{\cal W}$,

\begin{eqnarray}
\H_{D}'= & \frac{1}{4} \sum_{A\neq B} D^{AB} \left[ 2 T_{10}^{A} T_{10}^{B}+\frac{1}{2} \left(T_{11}^{A}T_{1-1}^{B} + T_{1-1}^{A}T_{11}^{B}\right) \right. \non\\
	 &+ 4 \left.\left(T_{21}^{A}T_{2-1}^{B} + T_{2-1}^{A}T_{21}^{B}\right) \right] 
\label{HinterZ}
\end{eqnarray}
where, $A$ and $B$ represent the pairs of protons of different water molecules and $D^{AB}$  describes the averaged intermolecular dipole-dipole coupling of protons of pair A with those of B. $T_{10}$, $T_{1\pm1}$ and $T_{2\pm1}$ are normalized spherical tensor operators of rank one and two, and orders zero and $\pm$1 , respectively. Notice that the tensor rank defines the spin number, while the order characterizes the coherence number.

The similarity transformation that rotates to the X basis, in terms of the spin operators is
\be
R=e^{-i \frac{\pi}{2} I_y} e^{-i \frac{\pi}{2} I_z}
\ee
 and the rotated Hamiltonian is
\begin{eqnarray} \label{HinterX}
\H_D^{\vert x\rangle} &=&R^{\dag}\H_{D}'R \non\\
&=& \frac{1}{4}\sum_{A\neq B}D^{AB}\left[- \frac{1}{2} \left( T_{10}^{A}T_{10}^{B}\right)^{\vert x\rangle} \right.\non\\
&-& \frac{5}{4}\left( T_{11}^{A}T_{11}^{B} +T_{1-1}^{A}T_{1-1}^{B} \right)^{\vert x\rangle} 
-\frac{3}{4}\left( T_{11}^{A}T_{1-1}^{B} +T_{1-1}^{A}T_{11}^{B} \right)^{\vert x\rangle} \non\\  
&+& 2\left( T_{21}^{A}T_{21}^{B} +T_{2-1}^{A}T_{21}^{B} +T_{21}^{A}T_{2-1}^{B}+T_{2-1}^{A}T_{2-1}^{B} \right)^{\vert x\rangle} \non\\
&+&\left. 2\left(T_{22}^{A} T_{22}^{B}+T_{2-2}^{A}T_{2-2}^{B}-T_{2-2}^{A} T_{22}^{B}-T_{22}^{A} T_{2-2}^{B}\right)^{\vert x\rangle}   \right]
\end{eqnarray}
where the symbol $\vert x\rangle$ labels operators in the X basis.              
It is clear that $\H_D^{\vert x\rangle}$, obtained after appropiate truncation of the dipolar interaction, when expressed on the X basis contains two-spin  and four-spin order terms, and that only even order coherences up to $\pm 4$ arise.
}
\section{Captions to figures}
Figure 1: Dipolar signals after the Jeener-Broekaert experiment and Zeeman signals (dotted) in (a) nematic 5CB  and (b) powder adamantane,  at 7 T and 301 K.  In (a):  dipolar \St $\;$ (solid) corresponds to a preparation time $t_{12}$=28 $\mu$s  and dipolar \We $\;$ (dashed) to  $t_{12}$=70 $\mu$s. In (b): several dipolar signals corresponding to different $t_{12}$. Vertical lines indicate the times of maximum and zero \St-order.\\
  
Figure 2: Radio frequency pulse sequence used to encode coherences of the spin states at time $\tau$ after the Jeener-Broekaert pulse pair, on the Z and X basis. Phases $\phi$ and $\eta$ were incremented from zero to 2$\pi$ in steps of $\pi$/8 to encode up to eight quantum coherence in both bases.  
$\epsilon$ is set short enough to minimize evolution during this period. Setting $\delta$ before the last 45$^o$ readout pulse sufficiently long, undesired transient signals are allowed to decay. The signal corresponds to 5CB prepared in a state of \St $\;$ dipolar order, $\tau$=300 $\mu$s, $\epsilon$=1 $\mu$s and $\delta$=1 ms.\\

Figure 3: X and Z basis correlated coherence numbers at short evolution times $\tau$ after the Jeener-Broekaert pulse pair before the establishment of the quasi-equilibrium states in (a) powder adamantane, (b) nematic 5CB prepared in the \St $\;$ condition and (c) 5CB prepared in the \We $\;$ condition.\\

Figure 4: Evolution of the coherence amplitudes for times $\tau$ after the Jeener-Broekaert pulse pair  projected on Z and X basis. (a) Powder adamantane, (b) 5CB prepared in the \St $\;$ condition, (c) 5CB in the \We $\;$ condition, (d) same as (c) in the range 0 to 5ms.\\

 Figure 5: Z and X basis projections of the coherence spectra (a), (b) at $\tau$=300 $\mu s$, (c) at $\tau$=2 ms. Powder adamantane (a) and 5CB in the \St $\;$ state (b) have similar shapes, while 5CB in the \We $\;$ state (c) show a completely different composition.\\

Figure 6: Relaxation of the multiple quantum coherences encoded on the X basis, starting from  dipolar order in (a) powder adamantane ($T_{1D}=665\pm$30 ms), (b) 5CB \St $\;$ condition ($T_{\cal S}=310\pm$13 ms) and (c) 5CB \We $\;$ condition ($T_{\cal W}=126\pm$15 ms), at 7T and 301K. Inset in (b) shows relaxation of coherence $x=\pm 4$ observed in 5CB under the \St $\;$ condition.\\

Figure 7: Dipolar relaxation time vs coherence numbers in powder adamantane (solid diamond), 5CB \St $\;$ (solid circle) and \We $\;$ (open circle) order at 7T and 301K.
\newpage
\begin{figure}[!hp]
\vspace{20cm}
\hspace{-11cm}
\special{eps: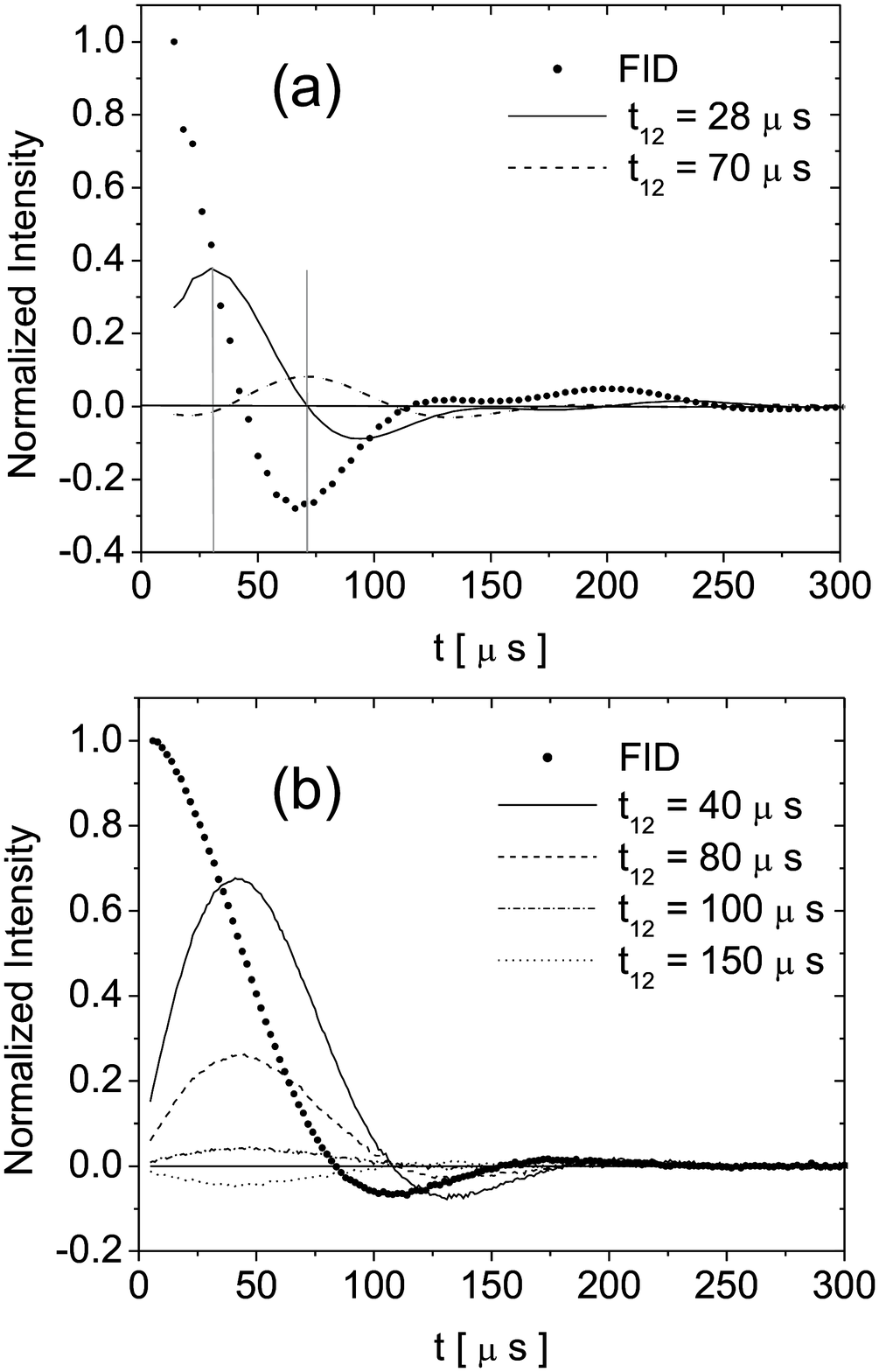 x=10.23cm y=15.48cm}
\caption{}
\label{uno}
\end{figure}  

\newpage
\begin{figure}[h*]
\vspace{10 cm}
\hspace{-9cm}
\special{eps: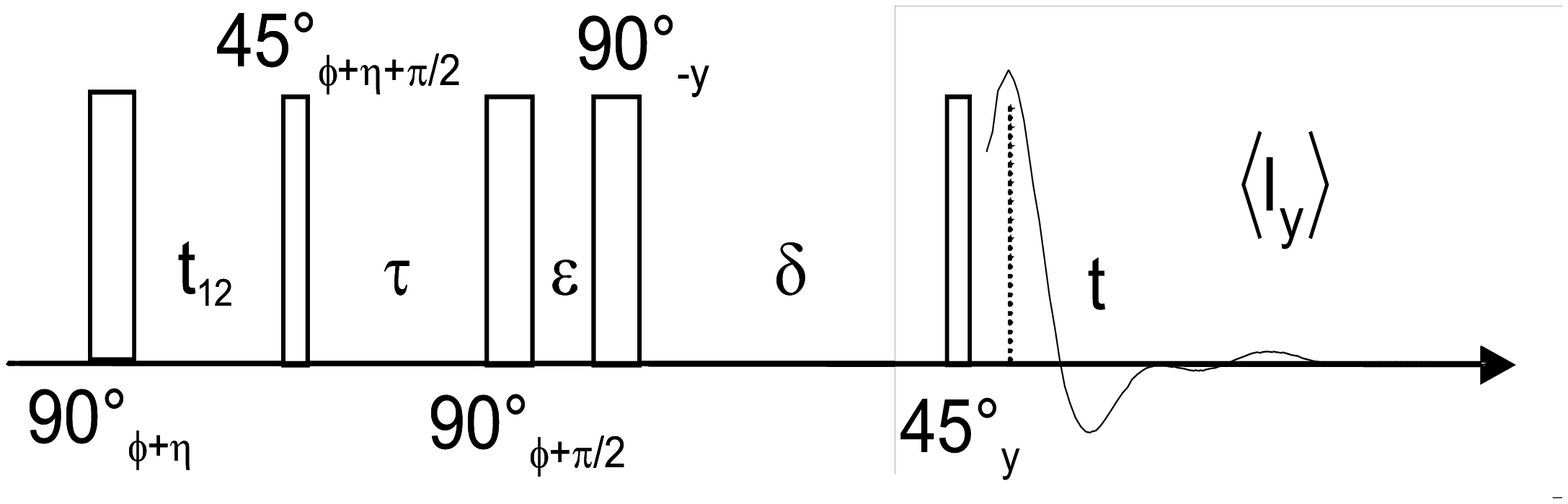 x=8.47cm y=2.86cm}
\caption{}
\label{dos}
\end{figure}  

\begin{figure}[h*]
\vspace{15cm}
\hspace{-9cm}
\special{eps: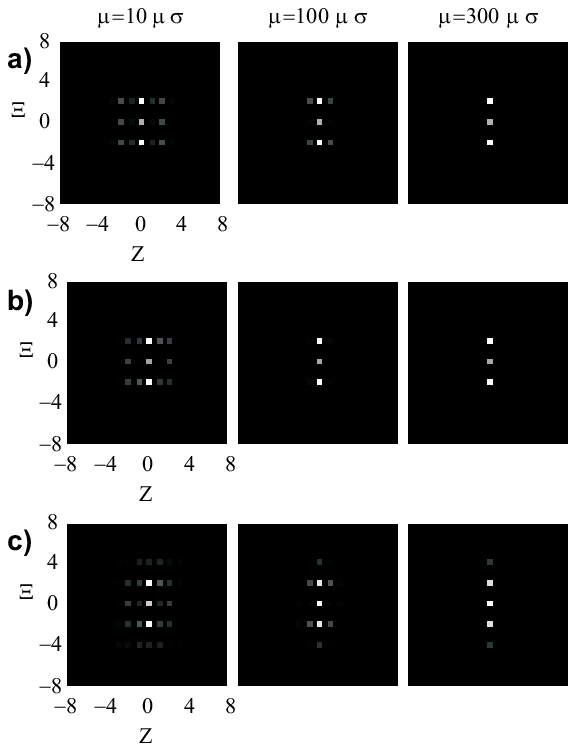 x=8cm y=10.81cm}
\caption{}
\label{tres}
\end{figure}  

\begin{figure}[h*]
\vspace{20cm}
\hspace{-10cm}
\special{eps: 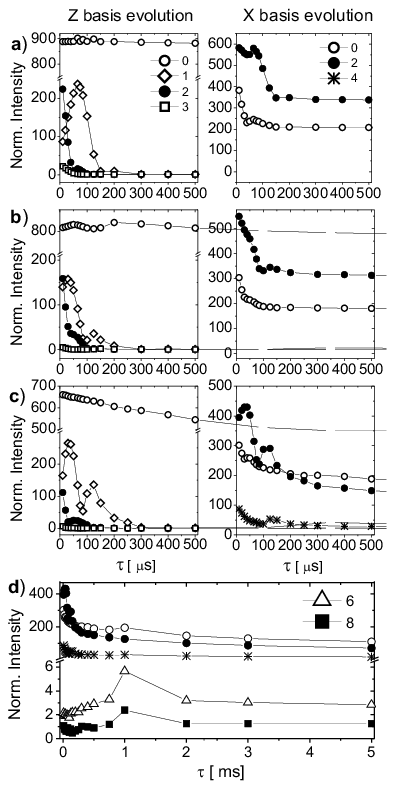 x=8.5cm y=17.54cm} 
\caption{}
\label{cuatro}
\end{figure}

\begin{figure}[h*]
\vspace{20cm}
\hspace{-11cm}
\special{eps: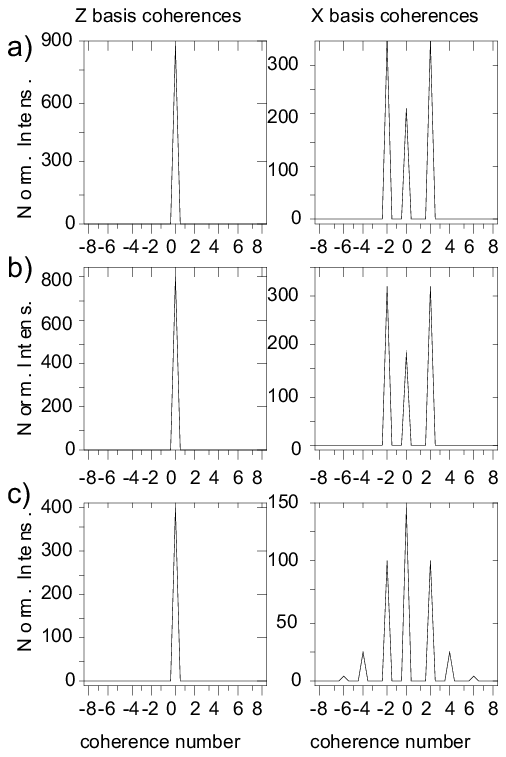 x=10cm y=16.25cm}
\caption{}
\label{cinco}
\end{figure} 

\begin{figure}[h*]
\vspace{18cm}
\hspace{-11cm}
\special{wmf: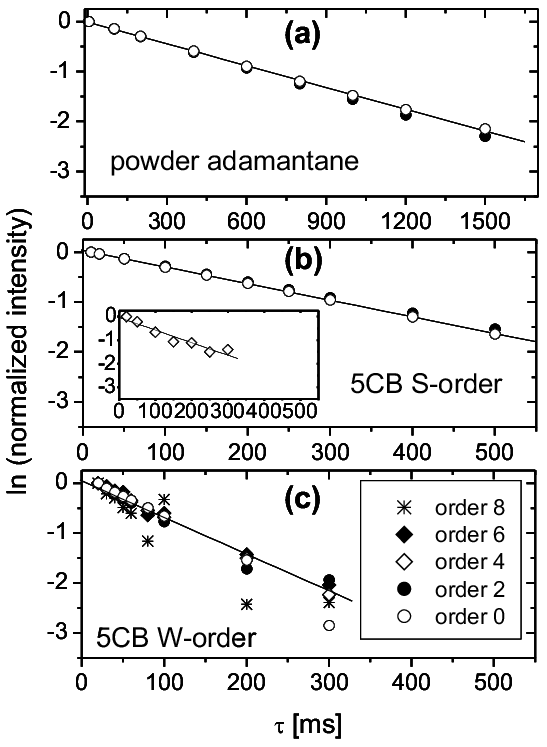 x=10.15cm y=13.38cm}
\caption{}
\label{seis}
\end{figure}

\begin{figure}[h*]
\vspace{15cm}
\hspace{-10cm}
\special{wmf: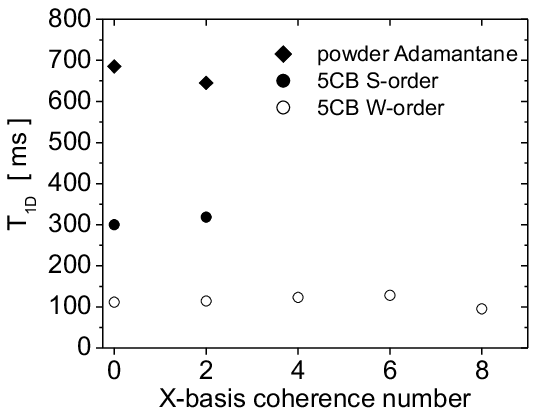 x=9.69cm y=7.66cm}
\caption{}
\label{siete}
\end{figure}

\end{document}